# The Governance of Risks in Ridesharing: A Revelatory Case from Singapore

By

**Yanwei Li, Araz Taeihagh\* and Martin de Jong**

\* Correspondence: Araz Taeihagh, Lee Kuan Yew School of Public Policy, National University of Singapore, 469B Bukit Timah Road, Li Ka Shing Building, Singapore 259771, Singapore spparaz@nus.edu.sg; Tel: +65-6601-5254

# The Governance of Risks in Ridesharing: A Revelatory Case from Singapore

Yanwei Li [1], Araz Taeihagh [2,*] and Martin de Jong [3,4]

[1] Nanjing Normal University, China

[2] Lee Kuan Yew School of Public Policy, National University of Singapore, Singapore

[3] Delft University of Technology, The Netherlands; [4] Fudan University, China

* Correspondence: Araz Taeihagh, Lee Kuan Yew School of Public Policy, National University of Singapore, 469B Bukit Timah Road, Li Ka Shing Building, Singapore 259771, Singapore spparaz@nus.edu.sg; Tel.: +65-66015254

**Abstract -** Recently we have witnessed the worldwide adoption of many different types of innovative technologies, such as crowdsourcing, ridesharing, open and big data, aiming at delivering public services more efficiently and effectively. Among them, ridesharing has received substantial attention from decision-makers around the world. Because of the multitude of currently understood or potentially unknown risks associated with ridesharing (unemployment, insurance, information privacy, and environmental risk), governments in different countries apply different strategies to address such risks. Some governments prohibit the adoption of ridesharing altogether, while other governments promote it. In this article, we address the question of how risks involved in ridesharing are governed over time. We present an in-depth single case study on Singapore and examine how the Singaporean government has addressed risks in ridesharing over time. The Singaporean government has a strong ambition to become an innovation hub, and many innovative technologies have been adopted and promoted to that end. At the same time, decision-makers in Singapore are reputed for their proactive style of social governance. The example of Singapore can be regarded as a revelatory case study, helping us further to explore governance practices in other countries.

## 1. Introduction

In past decades, we have witnessed the worldwide adoption of many different types of innovative technologies aiming at delivering more efficient and effective public services [1]. The sharing economy is such an example [2]. It is also called the 'collaborative economy', 'collaborative consumption', or 'peer-production economy', and is defined as any marketplace that brings individuals together to share or exchange otherwise underutilised assets [3,4]. The sharing economy allows customers to interact with service providers through innovative technologies. Many sharing industries, like Uber, BlaBlaCar, AirBnB, TaskRabbit, and Grab, have established themselves in many different fields, such as transportation, hospitality, and consumer goods. In the United States, millions of citizens rent rooms, cars, and even power tools from total strangers. The sharing economy in recent years has received substantial attention because its business model has disrupted many traditional industries by changing the way in which supply and demand are matched in real-time [5].

In the transportation domain, the uncertainty of energy cost, together with the pressure to increase energy efficiency and reduce carbon emission, has fuelled interest in seeking alternatives to private vehicle use [6,7]. Ridesharing is such an alternative [8], and in North America, it can be traced back to "car clubs" or "carsharing clubs" during World War II [9]. Traditional carsharing (like carpooling) moves people by using existing infrastructure and vehicles more efficiently. However, it is limited in responding to flexible commuting options [10]. From the 1990s, some ridematching services have integrated the use of mobile phones, social networking, and the internet with the service they offer. A new generation of ridematching platforms also known as 'dynamic ridesharing' has been developed since 2004 [11]. This new generation of ridesharing allows individuals to access the benefits of private cars without the obligations of car ownership [9]. Some ridesourcing companies, such as Uber and Grab, are examples of this new generation of ridesharing and the focus of our work. Throughout this manuscript when we refer to ridesharing, we are referring to the developments of this new generation of ridesharing companies in Singapore and the service they provide. It increases occupancy rates and the efficiency of urban transportation systems [12]. It offers commuters a more comfortable and time-efficient travel option and allows owners of vehicles to recover some of the journey costs [13]. Furthermore, ridesharing is helpful in reducing congestion, pollution, emission, and demand for parking infrastructure [14–19]. Ridesharing may also play an important role in redistributing income by providing poor residents the opportunity to access cheap transport services or secondary jobs [13]. The democratic and policy implications have also been discussed by public policy and governance scholars; they argue that ridesharing enables people to communicate with one another and increase social capital, and it is regarded as a new form of civic engagement in policy fields, which provides opportunities for citizens to engage directly in developing and implementing solutions and services (for example, resolving congestion problem and reducing emission through increasing utilisation rates) [20]. In short, ridesharing is viewed by some as highly beneficial for both individuals and society as a whole.

Ridesharing provides productivity gains by increasing the rate of asset utilisation and reduction of transaction costs and regulatory overheads [21]. However, ridesharing is disrupting existing markets. It is now facing pressure and pushback from the taxi industry everywhere. It may also result in concerns about the safety and privacy of customers, unfair competition, and service quality [22–24].

As some regulation and governance scholars have recognised, innovation and risks are inextricably interrelated, since innovation is uncertain both in terms of process and outcome [1,6,25]. Little is known about how to govern the risks involved in innovation (specifically ridesharing in this article). This begs the following research question: how can the risks involved in ridesharing be governed? Different cities/governments have adopted different strategies to regulate ridesharing. Some countries, like Germany, have banned the use of ridesharing services, such as Uber. Other places, such as New York City, tried to ban it but decided to hold back in response to commuters' demands [26]. In short, it seems that governments in different jurisdictions have adopted different strategies in addressing the risks in relation to the adoption of ridesharing. We will answer our research question through an in-depth case study. Singapore is our chosen case, and we will intensively examine how the risks involved in ridesharing are governed by the Singaporean government.

In Section 2, we present our theoretical framework. Following this, Section 3 lays out the method used and the case description. In Section 4, we identify the strategies applied by Singapore's government to address the risks associated with ridesharing. The final section consists of a discussion and conclusion.

## 2. Analytical Framework

This section is structured in three parts. The characteristics of ridesharing are summarised in Section 2.1. The risks associated with ridesharing are elaborated in Section 2.2. Five governance strategies in addressing these risks are elaborated in Section 2.3.

*2.1. Ridesharing*

Three different types of ridesharing have been identified based on the relationships between participants: acquaintance-based, organisation-based, and ad-hoc ridesharing [21]. Acquaintance-based ridesharing is generally among families, friends, and co-workers. Organisation-based ridesharing demands participants to receive the services by formal membership [22]. Ad-hoc ridesharing is achieved through various computerised ridematching platforms. In this study, we focus on ad-hoc ridesharing, also called "real-time ridesharing" or "instant ridesharing" [11]. The concept of ridesharing as discussed in the remainder of this article refers to ad-hoc ridesharing.

Agatz et al. identified a number of key features of ridesharing: dynamic, independent, cost-sharing, non-recurrent trips, and automated matching [11]. The details of these features are shown in Table 1.

**Table 1.** The five features of ridesharing (based on [11]).

| Features | Details |
| --- | --- |
| Dynamic | Ridesharing allows drivers with spare seats to establish links with users who want to have an on-demand ridesharing. |
| Independent | Drivers are not employees of ridesharing companies. |
| Cost-sharing | The costs during the trips are shared by participants. This is cost-effectiveness for users. However, some ridesharing drivers are profit-seeking. |
| Non-recurring trips | Trips of ridesharing are non-recurring, implying that ridesharing consists of one-way trips. |
| Automated matching | A system helps riders and drivers match each other instantly. |

*2.2. Risk in the Adoption of Innovative Technologies*

Adoption of many technological innovations, such as big data, waste incineration power plants, crowdsourcing, autonomous vehicles (AVs), the internet of things, and block-chain technology can cause new problems, since they often have unintended consequences and create new, previously inconceivable risks [27,28]. As a result of this, social acceptance of such innovative projects may be low [1,29].

Risk and uncertainty are two highly related concepts. The former implies that probabilities of events and their possible outcomes are known, whereas the latter means that we know some events might occur, but we do not know their probabilities [30–32]. Different types of risk are present in the adoption of innovative technologies, for example, market risk, political risk, technological risk, finance risk, environmental risk, organisational risk, social risk, and turbulence risk [33–38]. In this article, we focus our attention on technological risk, which is defined as the potentially negative social, economic, and physical consequences related to citizens' concerns in adopting innovative technologies [39]. Our interest is in identifying citizens' concerns since citizens are direct users of innovative technologies and are likely to be negatively influenced by them.

*2.3. The Governance of Risks in Ridesharing*

Some researchers have acknowledged that the adoption of innovative technologies will inevitably result in some unanticipated risks [40,41]. Often these unknown risks tend to cause substantial losses for the involved actors [42]. As such, the governance of these unknown risks—which are inherently involved in the adoption of innovative technologies—is of crucial importance.

The issue of the governance of technological risks has received scholarly attention from different perspectives, such as the field of governance and public policy [1,6,43–47], planning [48], risk management and governance [49–51], science and technological policy [52,53], complexity science [54–56], and organisational sociology [42,57]. They provide different answers to the question of how technological risks associated with the adoption of innovative technologies are governed, and identify some governance strategies, such as resistance, prevention, resilience, robustness, antifragility, adaptation, risk assessment, deliberation and negotiation, public participation, and learning by doing [53,54,56]. After a literature review, we identify five governance strategies that can be adopted by decision-makers in governing risks involved in the adoption of innovative technologies: no response, prevention-oriented strategy, control-oriented strategy, toleration-oriented strategy and adaptation-oriented strategy [1,32,45,49,50,52,55–57]. These strategies are elaborated below (for a review of the various strategies to govern risks of innovative technologies see [58]):

1. No response: no specific actions are taken by decision-makers to address risks. This might happen because decision-makers are not familiar with the potential consequences of innovative technologies [31]. If the nature of the innovative technology is unknown, decision-makers tend to put off decisions [55]. When ridesharing services were launched in the summer of 2012, some local governments did not know whether these new services should be established as peer-to-peer taxi services, ridesharing, or for-hire vehicle services [18]. No response also means that decision-makers do not develop regulatory frameworks to cope with potential dangers and threats.

2. Prevention-oriented strategy: decision-makers take preventive actions with the aim of eliminating risk in adopting innovative technologies [53,59]. We, for instance, can build a wall to prevent the invasion by enemies. In terms of the governance of risks in adopting innovative technologies, this strategy implies that decision-makers may prohibit the adoption of innovative technologies to avoid the existence of risk. The prevention-oriented strategy is an appropriate option for situations which are highly predictable, but it is slow in its responsiveness, and it might result in system paralysis [60].

3. Control-oriented strategy: this strategy favours traditional risk assessment. It assumes that scientific knowledge is helpful in narrowing the supposed uncertainties [61], and a centralised form of governance style is necessary [45]. In this strategy efforts to eliminate risk are considered sub-optimal and achieving an acceptable level of risk through rational risk assessment is favoured [62,63]. This partially corresponds with our understanding of the regulatory state, which emphasises that decision-makers can regulate risk through formal rules or policies [64]. An example of this strategy is that policy makers apply existing policies to regulate innovative technologies with the aim of controlling the involved risk [65].

4. Toleration-oriented strategy: the main aim of this strategy is risk tolerance [66]. It means that some actions are taken by decision-makers to prepare the systems or organisations to perform well in a constantly changing environment. This strategy partially corresponds to the understanding of resilience formed by many scholars, which means that a system or organisation can survive a wide range of uncertainties [67,68]. Developing alternatives is the first option under this strategy. One instance is when governments prepare several different sources of energy for potentially expected events [53]. A second option under this strategy can be a policy change or reform [55].

5. Adaptation-oriented strategy: decision-makers attempt to improve the adaptive capacities of the regulated system or organisation. It corresponds with the idea of adaptive resilience identified by some researchers [43,44]. Many different tactics match this strategy, such as learning by doing, public participation, forward-looking planning, co-deciding, and negotiation [45,48,56,57,67–69]. One example is the establishment of an independent review committee facilitating information access to the public [70]. Another example is roundtable discussions involving representatives of all potential stakeholders that are organised by policy makers in the governance of risk in mobile telephony. No clear rules and specific procedures are established, and a vague list of issues to be discussed is developed. The chair of these discussions allows all participants to voice their opinions and concerns [49].

These different government strategies function in this article as a heuristic tool to facilitate our analysis of the governance of risk associated with the adoption of ridesharing in Singapore.

## 3. Method and Case Description

The aim of this article is to answer the question of how technological risk associated with the adoption of ridesharing is governed. The case study approach is an appropriate strategy for answering how-oriented research questions [71]. Three reasons justify our choice of Singapore. First, Singapore is one of the world's leading innovation hubs [72]. Many different types of innovative technologies are adopted there, and the Singapore government has accumulated substantial experience in governing risks associated with innovative technologies. Second, the Singapore government has a reputation for proactive governance. Findings in this article can provide practical suggestions and implications for decision-makers in other countries. Third, our location made it most practical for us to conduct interviews with stakeholders involved in the governance of risks in the adoption of ridesharing in Singapore. We view the Singapore case as revelatory because few studies have been conducted to research comprehensively how risks associated with ridesharing are governed. Through interviews with government officials, researchers of transportation policy and the sharing economy, a social media writer and taxi driver, and a newspaper correspondent. Data collection and interviews were carried out by the first author during 2017. Respondent Number 1 was a Newspaper correspondent, Respondents 2 and 3 were sharing economy researchers, respondents 4 and 5 were Government officials, respondent 6 was a Transportation policy researcher, and respondent 7 was a social media writer and taxi driver. We traced the process regarding the governance of risks in the adoption of ridesharing, and identified the risks involved and strategies applied by the Singapore government in coping with risks in adopting ridesharing. Each interview lasted about one hour on average. The main comments made by respondents were recorded using notes. In addition, a large amount of secondary data was collected. We focused on the two most well-known mainstream media, Strait Times and Channel News Asia. We also collected information from the website of the Land Transport Authority (LTA) of Singapore and the discussions in the Parliament of Singapore. Based on this extensive body of secondary data along with the interviews conducted, the processes regarding the governance of risks associated with ridesharing can be summarised.

We have examined the governance of ridesharing from its initial introduction in Singapore in January 2013 to January 2018. Uber and Grab are the two dominant ridesharing companies in Singapore. They offer a range of services that can be booked by smartphone. The platforms set fares based on distance, location, and demand and the ridesharing companies take a 20 percent commission of the charged fares. In 2010, Uber (initially called UberCab) was launched in San Francisco and later in 2012 UberX a peer-to-peer ridesharing service was introduced. Uber did not own fleets of passenger cars, and they instead recruited individual car-owners and drivers. In September 2014, ridesharing services were provided by Uber in over 200 cities in 45 countries. Uber states that its vision is having fewer vehicles on the roads and allowing access to drivers within minutes and Users of Uber can "e-hail" different options ranging from taxi service to limousines, and rides with seats for people with special needs and children as well as amateur drivers and carpooling (through UberPool) [73].

Tan Hooi Ling and Anthony Tan founded Grab, which is mostly used by citizens in South-east Asia. Similar to Uber, Grab also expanded from car-hailing to GrabCar (ridesharing), GrabBike (motorcycle hailing), GrabExpress (delivery service), and GrabHitch (carpooling service). Grab has raised $700 million since 2012 and is considered one of the most successful start-ups in Southeast Asia (see https://www.grab.com/sg/about). The governance processes of ridesharing in Singapore can be categorised into five different phases thus far, and they are presented below.

*Phase 1: Hands-Off of Ridesharing (January 2013–September 2014)*

Uber started its trials in Singapore in January 2013 and officially launched its carsharing service in February 2013. GrabTaxi started in Malaysia in 2011 and launched in Singapore in October 2013. In March and September 2014, Uber launched UberX and Uber Taxi, respectively, in Singapore [74]. Initially, Uber and Grab were operating in Singapore unfettered. The ridesharing companies saw themselves as technology firms rather than transport providers [75]. These companies were viewed as a viable means of addressing problems with access to taxis during peak hours, and they were described as technology companies that provided riders and drivers matching services in Singapore.

*Phase 2: Regulating Ridesharing (November 2014–July 2015)*

With the popularity of ridesharing, taxi drivers in Singapore started to view ridesharing companies as competitors [76]. They argued that the private-car drivers should be subject to various regulations that applied to taxis, and at the same time changed their business practices to allow taxi bookings using ridesharing platforms. Taxi drivers voiced their disapproval of the wide application of ridesharing in Singapore. However, Singapore government has established the "Smart Nation" programme as a national vision and is promoting adoption of new technologies, implying that it would be counter to the Smart Nation vision to ban the use of the ridesharing apps such as Uber in Singapore. Recognising the legitimate concerns of the taxi drivers, Singapore government sought to increase the fairness for all players in the marketplace.

In November 2014, LTA started to regulate third-party ridesharing companies and issued five regulations to be put into effect from summer 2015. The details of these regulations are shown in Table 2 [77]. Moreover, the LTA stated that it would consult the relevant actors such as the National Taxi Association (NTA), taxi companies, third-party taxi booking services, and commuters about regulation of third-party booking services.

Table 2. Regulations released in 2014 about third-party ridesharing companies.

| | Details |
|---|---|
| 1. | Registering with the LTA: the third-party ridesharing companies are required to register with the LTA and if successful in their application will be authorised to carry out their service for a period of three years; |
| 2. | Only dispatching licensed drivers with taxi vocational licences to ensure riders are served by drivers that are operating legally in Singapore; |
| 3. | Making all fees (flag down and booking fees) and surcharges (location, peak period) and rates (per time and distance) available to commuters before the service commences; |
| 4. | Not requiring consumers to specify their destinations before making a booking; |
| 5. | Providing basic customer support (such as providing avenues for feedback or complaints and lost-and-found services). |

On 11 May 2015, Parliament approved the bill, the Third-Party Taxi Booking Service Providers Act. This act demanded all third-party taxi booking companies that had more than 20 participating taxis should register their service with LTA and receive their certificate from LTA before they can start their operation [78]. In addition, service providers were required to provide the live data on their booking to LTA [79]. LTA allowed existing service providers that did not need a certificate to operate before the commencement of the law to continue their operation if they registered with the LTA before 1 December 2015 and until their application was processed [80]. The application of PAIR Taxi was rejected because it did not adhere to the new fare charging framework. Under the new regulations operating without a certificate of registration with LTA could cause a fine of up to $10,000 or imprisonment up to six months or both [80]. One 8 July 2015, Uber increased its fare during evening's SMRT train disruption. This led many users of the booking app felt unhappy [81]. One week later, on 17 July, LTA stated that surge pricing was used only for chauffeured vehicle booking services and not its taxi booking service [82].

*Phase 3: Collecting Opinions of the Stakeholders Involved in Ridesharing and the Implementation of the New Regulation (October 2015–December 2015)*

In October 2015, Senior Minister of State for Transport led an industry review. Various parties, including commuters, taxi drivers, taxi companies, ridesharing companies, and private hire car drivers were consulted. The NTA called for 'fair competition' and stated that commuter safety should be protected through requiring private-hiring car drivers to pass the same checks as regular taxi drivers. On 13 October 2015, the General Insurance Association of Singapore advised private owners who are providing ridesharing service using their vehicles commercially to acquire appropriate insurance coverage as normal insurance coverage does not cover such activities [83]. On 11 November 2015, Grab announced plans to launch GrabHitch, a carpooling service that provides a low-cost, door-to-door transport service, with prices closer to public transport [84].

Five days later, on 16 November 2015, the NTA, along with ten taxi drivers, met with Mr. Ng to share concerns and recommendations [85]. On 28 January 2016, GrabTaxi announced that it would combine its GrabCar and GrabTaxi services (and others) under a new parent company Grab [86].

*Phase 4: The Reforms of the Current Regulation Framework (April 2016–June 2016)*

On 12 April 2016, a new licensing framework, Private Hire Car Driver Vocational Licensing (PDVL) framework, was released that would come into effect in the first half of 2017 (see Table 3) [87]. This regulation required all drivers of the private hire car to have a background check and participate in a 10 h training course and pass the necessary tests. In addition, the existing Taxi Driver Vocational Licence (TDVL) was updated. The duration of the courses that the taxi drivers needed to attend was shortened from six- to nine-hour to between three- to five-hour and in addition the course now included training for using tools such as Global Positioning System (GPS) [87]. On 28 April 2016, Strides Transportation, launched a variety of ridesharing services and signed a one-year contract with Grab to allow its over 200 drivers to use the ridematching service offered by Grab [88]. On 13 June 2016, Grab stated its first cross-border carpooling which allows the commuters to share their ridesharing between Johor Bahru and Singapore [89]. One week later, on 20 June 2016, LTA indicated that it deemed the service model not compatible with regulations in Singapore and had informed the company about it [90]. As a result, the ridesharing service offered by Grab between Singapore and Johor Bahru, that had started on 20 June was changed to a free three-week trial [91].

*Phase 5: Further Levelling the Playing Field and Consolidation (August 2016–January 2018)*

On 21 August 2016, Prime Minister Lee Hsien Loong stated that the competition between taxis and ride-hailing services was still not quite level and that this issue would be further examined [92]. In September 2016, a new partnership between Trans-cab and Grab was announced to bring all of the Trans-cab fleet to the Grab platform [93]. Also, in the same month, Singapore government stated that it would review its Taxi Availability (TA) framework. The TA standards were established in January 2013. They required taxis drivers to work a certain minimum daily mileage and during the peak hours. In contrast, private-hiring cars did not have to adhere to these requirements. LTA claimed that it would monitor the situation carefully and guarantee the needs of the users and the welfare of taxi drivers [94].

**Table 3.** The Private Hire Car Driver Vocational Licensing (PDVL) framework and Taxi Driver Vocational Licensing [95].

| Screening | All applicants should undergo a medical examination and background checks. |
|---|---|
| Eligibility | Only Individuals that are employed by a limousine company or are registered as owners of such a company can provide chauffeured services. |
| Training | All applicants must successfully pass a PDVL course and participate in refresher course every six years. PDVL holders with no demerit points, and who have been active drivers, will be exempt from the refresher course. |
| Disciplinary measures | PDVL licensees would be subject to the Vocational Licence Points System (VLPS). |

On 23 September 2016, nuTonomy started testing two driverless cars it developed in a partnership with Grab to allow its users to try them out for free [96]. On 25 September 2016, in an accident involving a private hire car under Uber, a woman was killed [97]. On 17 December 2016, LTA released its review report about TA, and it stated that the "percentage of taxis with minimum daily mileage of 250 km" requirements would be removed [98]. In January 2017, a new law proposed to give a new set of powers to authorities to crack down on ridesharing provides in case their drivers operate without proper insurance or licences [99]. Moreover, in January 2017, LTA announced that ride-hailing services such as Uber and Grab required car seats for young children as under Road Traffic Act, carrying passengers under 1.35 m was illegal without a booster seats or child restraints, and these services, unlike taxi services that were public service vehicles, were not exempt from this ruling.

On 7 February, the Minister for Transport stated that it would create new rules to share data from the trials, set standards for the design of the AV equipment, and place time and space limits on AV trials [100], and The Road Traffic Act was amended [101]. The Road Traffic Act now recognised that in AV testing the vehicle is no longer in control of the human driver and exempted the AV system and its operator and people involved in the trials of AVs from provisions that held human driver responsible for the use of vehicle and recognised that the vehicle is now controlled by the AV system. To avoid stifling innovation a five-year regulatory sandbox was developed before enactment of legislations. In the meantime, AVs need to pass safety assessments and must have accident mitigation plans before testing can commence [100].

Furthermore, news came out that PDVL would be implemented from the second half of 2017 [102]. The drivers had to apply for the PDVL by 30 of June, after which they would be given up to a year to successfully complete the 10-h PDVL course, undergo medical and background checks and have proper insurance and be in possession of Class 3 driving licence for at least two years [99,103]. Operating a private hire car without a decal or vocational licence can result in a fine of up to $1000, jail up to three months, or both, in the first instance [104]. Later in March 2017, LTA opened the applications for PDVL and almost immediately afterwards, both Uber and Grab stated that they would pay the costs of the drivers to obtain their licences [103]. In June, over 90 percent of drivers of Uber had registered for the FastLane programme that was rolled out in March to facilitate the process of the drivers attaining the PDVL as it became a requirement for all private hire drivers [105].

Furthermore, in March 2017 new amendments were introduced in parliament to existing legislation in Singapore addressing cybercrime. The Computer Misuse and Cybersecurity Amendment were passed in May 2017 criminalising dealing and trading personal information such as credit card records even if the trader may not have intruded into a computer system for gaining access to the information, the amendment also includes crimes committed outside Singapore and on computers overseas if such an act causes significant risk of harm in Singapore [106]. In June 2017, Grab announced extending its group personal accident insurance to cover its carpool services aiming to complement the personal motor insurance purchased by their drivers and guarantee safe travel to its riders [107]. Moreover, a new Cybersecurity Act was scheduled to be introduced in the middle of 2017 to address any existing legislative gaps after public consolations. The bill was released in July 2017 giving Cyber Security Agency (CSA) power to investigate cyber security incidents while working with regulation in that sector. Consultations were held in July and August 2017, and it was reported that 92 entities provided feedback which was mostly positive and resulted in minor edits to the proposed bill which is scheduled for introduction in 2018 [108].

In early August 2017, the Wall Street Journal learned that Uber had knowingly bought more than 1000 defective Honda Vezels and through their Lion City Rentals, rented them to drivers while knowing they had been recalled and one had caught fire in January 2017 [109]. By 17 August just two weeks after the news broke out, according to LTA, Uber had recalled the entire 1220 defective Hondas and fixed them, and LTA stated it had been monitoring the recall process [110]. In August 2017, the largest taxi company in Singapore, ComfortDelGro, signed a letter with Uber for exclusive discussions on a potential alliance with Uber, which could made ComfortDelGro's 15,500 taxis available through Uber's app and given its in-house engineering and maintenance capabilities help forming a partnership in managing the 15,000 vehicles owned by Lion City Rental that is an Uber-owned company as well [111,112]. The deal would be beneficial for Uber as well since it would get access to a large fleet of vehicles without the need for subsidise or funds [109]. It is worth mentioning that since March 2017 the other five taxi companies in Singapore (SMRT, Trans-Cab, Premier, Prime, and HDT Singapore Taxi) already had made similar arrangements with Grab and offered dynamic fare pricing and in October 2017 Grab announced an exclusive partnership with SMRT that provided Grab drivers exclusive access to SMRT's maintenance and engineering services [113].

In November 2017, news broke out that Uber had a major security breach in 2016 which compromised personal details of 57 million accounts worldwide and had paid ransom to hackers [114]. Following the news of the incident, various experts highlighted that laws must be tightened and commercial entities should inform consumers and authorities when such breaches occur. As mentioned earlier, major changes to Personal Data Protection Act (PDPA) were proposed in Singapore in July 2017 which included requiring the organisations that suffer from a breach to notify consumers as soon as they become aware of it and in case the data of more than 500 are involved Personal Data Protection Commission (PDPC) must also be notified within 72 h and in case critical infrastructure is involved Singapore's Cyber Security Agency (CSA) must be informed based on a new proposed cyber security bill which is expected to be debated in 2018.

Later, in December 2017, news broke out that the personal information of 380,000 of accounts in Singapore were compromised as part of the breach in 2016, and subsequently the PDPC stated that it had started investigation whether Uber was in breach of any laws in Singapore and LTA made similar statement, expecting full transparency and cooperation with regulators investigating the breach [115]. In the same month, ComfortDelGro announced its intentions to acquire 51 percent stake of the Lion City Holdings valued at $642 million, subject to regulatory approval and Uber and ComfortDelGro's applications will be integrated under "uberFLASH" and offer dynamic fare pricing if booked through Uber app similar to the Grab offering since March 2017 [113]. Given the implementation of PDVL earlier in July 2017, this deal would also enable Uber to increase its supply of drivers which through the implementation of PDVL might become harder to increase or maintain in future [116]. Following the announcement of the proposed Uber-ComfortDelGro collaboration The Competition Commission of Singapore (CCS), stated that due to competitive nature of the rental car market it would seek public consultation and feedback on the alliance starting from 21 of December 2017 to 8 of January 2018 [117]. While still under review, with special permission uberFlash service started operation in January 2018 [118]. After this first round of consultation, CCS continued a caution approach and carried out a second round of consultation in January 2018 and stated if there are concerns it will continue and carry out

more exhaustive deliberations given that with the alliance between Uber and ComfortDelGro would create a 27,000-vehicle fleet which will be the largest in Singapore.

## 4. Case Analysis

In this section, we summarise the types of risk involved in the adoption of ridesharing, then analyse how the risks involved in ridesharing are governed in Singapore.

*4.1. Risk in Ridesharing*

Ridesharing may result in technological risk. Based on our interviews and secondary data, we identify five key risks. They are elaborated in detail below.

1. Privacy: ridesharing platforms collect sensitive information about their customers, such as telephone numbers, geolocation data [119], and credit card numbers. Some users of ridesharing have complained about the ridesharing company's inappropriate gathering or use of data. Some critics even label Uber as a big data company that is transforming its business and focusing on leveraging the wealth of information it gathers to provide new services and generate revenue by selling this data to others [120]. If the collected private data on journalists, elected officials, and venture capitalists were used improperly, the outcomes might be disastrous [23]. Although Uber has been working on improving its privacy policy, it is not yet certain whether its policy will be effective in the future, and the Uber data Breach in November 2017 confirmed the difficulty of this task. Respondents 4 and 5 told us that privacy is not a serious concern for citizens in Singapore because the PDPA was already passed in 2012 to protect the privacy of citizens. Moreover, major changes and additional requirements were introduced to PDPA in summer 2017, and additional cybersecurity measures are to be introduced in 2018 through Cybersecurity Act. Private companies will be fined if they use citizens' private information in any illegal form. Currently, in response to the breach of personal information of Singaporeans, empowered by the existing legislation, the PDPC is investigating Uber.
2. Safety: customer safety raises concerns for users of ridesharing services in many countries [11]. Customers may feel unsafe when in cars driven by strangers [22]. Ridesharing drivers are not professionally trained and licensed like their tax driver counterparts. The background checks for drivers of ridesharing are not as strict as they are for taxi drivers. Drivers of ridesharing do not need to conduct fingerprint scans, as taxi drivers do. Cases of molestation, misappropriation, use of criminal force, and driving under the influence thus far have been addressed in courts [121–123]. In addition, the vehicles used for ridesharing may not be as rigorously inspected. For example, Uber does not require regular vehicle inspections, and same levels of safety inspections that conventional taxies go through are not applied to its cars [124]. Respondent 6 has argued that the main motivation of private hire car drivers is to earn more money. Many drivers in Singapore buy second-hand cars to work for Uber and Grab, due to the high cost of buying new cars (Respondent 7). This raises some concerns, as these drivers mostly want to earn more money and might not prioritise safety (Respondents 6 and 7) and indeed the acknowledgment by Uber that it could have done more in the case of defective Honda Vezels suggests that this problem is not limited to drivers [110].

3. Influence on incumbent industries: ridesharing is disruptive, implying that it may result in unanticipated consequences for incumbent industries. In this case, the taxi drivers viewed ridesharing companies as their competitors, and taxi drivers claimed that the use of ridesharing platforms such as Uber and Grab has negatively influenced their businesses (Respondents 6 and 7). The taxi industry argues that ridesharing does not have to comply with pricing or consumer protections and therefore has an unfair advantage [70]. Uber, for instance, employs a 'surge price' when demand is high. On 8 July 2015, Uber increased its fare during that evening's SMRT train disruption. This led to great dissatisfaction among many users [81]. On 25 November 2015, UberX in Singapore fares were reportedly as much as 3.8 times higher during North-South Line train disruptions [125]. According to the LTA data, the average of unhired taxis went up to 5.9 percent in the first 11 months of 2016 from 4.2 in 2015 [126]. This shows that the adoption of ridesharing disrupted the taxi industry and had a negative influence on the lives of taxi drivers.

4. Liability: ridesharing as an innovation may result in the rise of non-professional and non-regulated workers. Many ridesharing drivers use rental cars (Respondents 6 and 7) and may therefore not be insured. When accidents occur, this may result in losses for drivers and passengers. Insurance-wise, an unresolved legal question is who is liable when in an accident when a ridesharing vehicle is involved (Respondents 4–6). Uber has denied liability for accidents that occur during the use of its service [97]. Uber's argument is that it only provides a platform that facilitates matching drivers with passengers thus bearing no legal responsibilities for property damage or injuries caused by the drivers. Some, however, argue that ridesharing companies enjoy profits but offload risks onto others [70]. In September 2016, an accident involving a ridesharing vehicle occurred in Singapore, and a 19-year old girl died [97]. Additional deaths have resulted from the negligence of Uber and Grab drivers following this initial accident [123,127,128]. Currently, there is no consensus regarding the question of how to deal with ridesharing cars when they are involved in accidents and courts are addressing the issues on a case-by-case basis (Respondents 4 and 5). The introduction of mandatory insurance through PDVL for private hire drivers and mandatory car seat for children demonstrates the efforts to resolve the liability concerns.

5. Automation: With the development of ridesharing in Singapore, automation has seen considerable progress (Respondents 1, 4, and 5). As it was mentioned earlier, in September 2016, nuTonomy tested two driverless cars it had developed and set up a partnership with Grab to allow its users to try them out for free. In December 2016, Grab announced an investment from Japanese automaker Honda as part of a deal to collaborate on a motorbike-hailing service. Grab and Honda would form a partnership to develop ridesharing technology [129]. It is certain that autonomous ridesharing will come about in the near future, meaning that users will be able to book an AV for their journeys. Automation of ridesharing potentially can drastically change transportation systems as it can reduce fatalities due to crashes, and can increase the mobility of the elderly and disabled as well as increase road capacity, lower emissions, and increase fuel savings [130]. It may, however, result in new challenges for decision-makers. For example, it may lead to unemployment among taxi and private hire car drivers due to decreased demands on cars with drivers. Moreover, new unintended consequences are surfacing, such as the exacerbation of an organ shortage due to the decreasing number of deaths because of driver errors in motor-vehicles accidents by 94% [131]. The General Insurance Association is engaging the LTA on the implications of AVs for insurance, and still gathering facts on these new developments [132]. In our interviews, Respondents 1, 4, and 5 argued that the automation of ridesharing is not a major concern. Automation is still at the pilot stage and has a long way to go before it is widely adopted (Respondent 1). Currently, the LTA has established a partnership with A*STAR to provide a technical platform to test the use of AVs. Specifically, demarcated routes in a business park (one-north) located in Queenstown have been identified to support the testing of this technique [133]. For instance, certain time and space limitations have been set on trials involving driverless vehicles. Moreover, design standards will be established for the equipment used in such vehicles, and the developers involved in these tests will be required to share data (Respondents 4 and 5).

These five concerns are the main technological risks associated with ridesharing. They are highly related to the concerns of users and require serious consideration by decision-makers. After identifying these concerns, the strategies applied by the Singapore government to cope with them are elaborated in the following section.

*4.2. The Governance of Risks in Ridesharing in Singapore*

In this case, five different types of governance strategies in coping with risks associated with ridesharing can be identified. They are: the no response strategy, the prevention-oriented strategy, the control-oriented strategy, the toleration-oriented strategy, and the adaptation-oriented strategy. These strategies are elaborated below.

1. No response: when Uber and Grab started their operations in phase one, Singapore's government regarded them as an innovative technology that can efficiently achieve an automated matching between drivers and users [75]. As such, they essentially promoted their wide adoption, and it is argued that ridesharing is helpful in resolving traffic congestion problems (Respondents 4 and 5). At the time, no framework was established to specifically regulate ridesharing, which indicates a no response strategy.

2. Prevention-oriented strategy: it was reported that the carpooling service offered by Grab would soon enable people to use it to share a ride across the border between Singapore and Johor Bahru in Malaysia in Phase Four. However, the LTA in Singapore soon informed the company that the service model did not comply with Singapore's regulations. It is likely that Malaysian regulations do not permit Singapore-registered cars to provide ridesharing services either without a public service vehicle licence. The Singapore government thus decided to prohibit the ridesharing service of Grab between Singapore and Johor Bahru with the aim of eliminating the existence of risk (Respondent 1). This indicates the application of a prevention-oriented strategy.

3. Control-oriented strategy: to manage the development of ridesharing in Singapore and collect information about a number of private hire ridesharing drivers, it is necessary to take regulatory actions (Respondent 7). In this case, Parliament approved the Third-party

Taxi Booking Service Providers Act. Under this law, all third-party providers of taxi booking services are required to register themselves with LTA and comply with its regulations (Respondents 4 and 5). The Singapore government also developed a PDVL framework for private hire car drivers, which comes into effect in the first half of 2017 (Respondents 4 and 5). In addition, since Singapore's government passed the PDPA in 2012, and its amendment in 2017, the use of personal data is already strictly regulated (Respondents 4 and 5). Another example is the regulation of driverless ridesharing in Singapore. Respondent 1 argued that driverless ridesharing is a highly complex issue, as it might result in many unanticipated consequences for society. In February 2017, it was reported that the LTA would set time and space limits for driverless vehicles and establish design standards for equipment in such vehicles [100]. This all indicates an attempt to regulate the functions of ridesharing in Singapore, which points to the adoption of a control-oriented strategy by the government.

4. Toleration-oriented strategy: the taxi industry in Singapore has been over-regulated in past years (Respondents 2 and 3). Singapore increasingly recognised that the playing field needs to be levelled further, and the government has attempted to deregulate the taxi industry and promote their competitiveness vis-a-vis the ridesharing services (Respondents 4 and 5). It updated the regulatory framework for taxi drivers, TA, and removed certain regulations, such as the minimum daily 250 km mileage requirement to be upheld by a percentage of taxies, and the percentage of taxis operating during the peak hours (shoulder peak periods requirement), with the aim of further levelling the playing field (Respondent 6). The creation of a regulatory sandbox for testing AVs and amendment of Road Traffic Act to allow exemptions for AV testing further indicates that the reforms of current policies indicate the emergence of a toleration-oriented strategy.

5. Adaptation-oriented strategy: Singapore's government established a committee to review the risks regarding ridesharing in Phase Three. Respondents 4 and 5 told us that the LTA organised several rounds of interactions among different actors, including government officials, managers of taxi companies, representatives of taxi drivers, and managers of private hire car companies. These different actors exchanged ideas about the governance of ridesharing in Singapore. In addition, some taxi drivers sent emails or letters to officials of the LTA and the Ministry of Transport (MOT). Their complaints about the adoption of the ridesharing were well received, and explanations about the rationale behind the adoption of specific regulations for ridesharing were provided to them (Respondents 4 and 5). This shows that the Singapore government has attempted to enhance mutual understanding between various actors with the aim of building consensus regarding the nature of the risks involved in ridesharing. This consensus-based approach indicates the emergence of an adaptation-oriented strategy.

Note that the governance practice might be complicated, particularly in the context of addressing the risks of innovative technologies. A single government might adopt several different strategies simultaneously. In this case, the Singapore government combined the use of the adaptation-oriented strategy with control- and toleration-oriented strategies. It collected opinions from different stakeholders involved in the governance of ridesharing (adaptation) and released the Third-party Taxi Booking Service Providers Act and the PDVL framework (control), before finally reforming the TA framework (toleration) to improve the competitiveness of taxicabs. In addition, the government rejected the application of PAIR Taxi because its fare model did not meet the fare charging regulations of LTA. The rejection of PAIR Taxi's application could be due to the combination of a prevention-oriented (avoiding risk) and a control-oriented (regulating potential dangers and threats) strategy. The combined use of different strategies by a decision-maker might occur, but this does not downplay the usefulness of our typology of government strategies because its main aim is to provide an analytical tool for researchers to categorise government actions in the governance of risks associated with innovative technologies.

## 5. Discussion

Our case study suggests that an adaptation-oriented approach is a comparatively better approach for decision-makers in addressing technological risks associated with the adoption of innovative technologies (including ridesharing). In this case, the Singapore government has learned to apply such a strategy in coping with risks in ridesharing. We nevertheless must acknowledge the particularities of the Singapore case. First, innovation has been established as a national strategy by the government [72]. Singapore has been widely regarded as one of the most innovative countries around the world. In the Global Innovation Index (GII), a well-known index to measure innovativeness of nations, Singapore ranked the sixth most innovation nation [134]. Because of this, many technology companies view Singapore as a suitable environment to experiment new technologies. Innovation is a key feature of Singapore, and the government tends to support the development and application of various innovative technologies in Singapore. As a result, Singapore government prefers reconciling innovative technologies with the incumbent industries. Second, the Singapore government is highly responsive to the demands of citizens. After the establishment of Singapore, the government has promised to provide its citizens a high standard of living. As Respondents 4 and 5 argued, Singapore is a city-state, and its government has a tradition of applying a proactive style in societal governance. In the governance of social housing, the Housing Development Board (HDB), for instance, has established rather strict rules in responding to citizen feedbacks [135]. Citizens in Singapore have many channels for providing feedback such as through hotlines, emails, and letters, and HDB officials regularly collect these feedbacks, which are then analysed by Quality Service Managers. These analyses function as an important basis for HDB to make decisions. Because of these smart institutional designs, Singapore government has always treated citizen viewpoints seriously. It is therefore likely that it may apply an adaptation-oriented strategy in coping with the risks involved in ridesharing. Furthermore, an important factor contributing to the success of this proactive adaptation-oriented approach is the presence of a high level of policy capacity. Policy capacity is defined as the ability to arrange and allocate required resources to make intelligent decisions and to set appropriate directions that are beneficial to the public [136,137]. Presence of high policy capacity in Singapore and proactive engagement of the government [138,139] resulted in course adjustments by the ride-hailing companies as demonstrated through rapid reaction to PDVL, provision of child-friendly seating options, a two-week response and addressing defective vehicles by Uber and cooperation with LTA, PDPC, and CCS. Moreover, given the size and population of Singapore, it is relatively easy for Singapore's government to gain accurate information about the wishes of its citizens, which allows it to adjust its policies contingently.

Our case study has two crucial implications for decision-makers in other countries regarding the governance of risks in ridesharing. First, an adaptive approach is a favourable approach for decision-makers to prepare them for unanticipated consequences. Ridesharing is transforming the traditional transportation landscape around the world, and it is neither wise to prohibit its adoption nor to let it go unregulated. The former might discourage innovation, while the latter has a substantially negative impact on the incumbent industry. Our case study about Singapore may provide blueprints for other countries in the governance of innovative technologies. Decision-makers should actively collect the opinions and perspectives of different stakeholders involved in ridesharing and create a consensus regarding what are the best strategies for governing the risks. Second, the governance model developed in Singapore cannot necessarily be directly copied by other countries. As this research suggests, decision-makers in different countries should take into account the unique characteristics of the system they operate in (such as the state of the transportation subsystem, the level of acceptance of ridesharing among citizens, the governance structure—for example, the positions of taxi associations—and pressures from the traditional incumbent industry) and avoid a one-size-fits-all approach. For instance, a city with a serious traffic congestion problem might approach this issue differently from a city without a traffic congestion problem. Decision-makers should think carefully about the implications of their unique situations and design appropriate strategies for better governing the risks of ridesharing.

## 6. Conclusions

In this article, we have reported an in-depth case study to present which types of technological risks concern Singapore's government and citizens and examined how the government has addressed these risks so far. We found that Singapore's government has identified five different types of technological risk: privacy, liability, automation, safety, and the impact on incumbent industries. To cope with them, Singapore's government has adopted five different strategies: no response, and prevention-, control-, toleration-, and adaptation-oriented responses.

A few studies have been conducted to explore how decision-makers govern innovative technologies [1,6]. This study is revelatory and provides insights for decision-makers in other countries on how to proactively address the risks of ridesharing. The wide access to smartphones and the internet makes it easier for citizens around the world to use ridesharing to meet their travel needs. Governments face difficulties in governing the risks associated with ridesharing, and with growing rates of adoption and the possibility of widespread automation of these services in the future, this issue is becoming a priority. Our case study showed a comparatively positive example regarding the governance of risks in the adoption of ridesharing. The Singapore government neither prohibits the development of ridesharing nor lets its development go completely free. Rather, it takes proactive measures to level the playing field to achieve both the wide application of this innovative technology and increase the competitiveness of the incumbent taxi industry. In this case, it seems that the Singapore government has applied an adaptation-oriented approach to address the technological risks associated with ridesharing, collecting the opinions, perspectives, and ideas of different stakeholders involved in its governance. It favours a bottom-up approach in resolving the complaints of the taxi drivers and attempting to have face-to-face negotiations with them to build consensus. Moreover, with the rapid technological advancement and development of new business models, Singapore's government is proactively monitoring the implementation of its policies on regulating ridesharing and making adjustments to address potential concerns.

**Author Contributions:** Y.L., A. T., and M. J. all have contributed significantly to the research design, data analysis and writing of the manuscript. Y. L. performed the interviews and carried out the data collection.

**Acknowledgments:** Araz Taeihagh is grateful for the support provided by the Lee Kuan Yew School of Public Policy, National University of Singapore through the Start-up Research Grant. Yanwei Li is grateful for the support provided by the Jiangsu Provincial Department of Education through the project Governing the sharing economy, grant no.: 111360B31701.

**Acronyms**

| Acronyms | Full Name |
|---|---|
| AV | Autonomous Vehicle |
| CCS | Competition Commission of Singapore |
| CSA | Cyber Security Agency |
| GPS | Global Positioning System |
| LTA | Land Transport Authority |
| NTA | National Taxi Association |
| PDPA | Personal Data Protection Act |
| PDPC | Personal Data Protection Commission |
| PDVL | Private Hire Car Driver Vocational Licensing |
| TA | Taxi Availability |
| VLPS | Vocational Licence Points System |

35 of 37